\documentclass[prb,superscriptaddress,showpacs,twocolumn,a4paper]{revtex4}
\usepackage[dvips]{graphicx}
\usepackage{amssymb}
\newcommand{\ped}[1]{\ensuremath{_{\rm #1}}}
\newcommand{\apex}[1]{\ensuremath{^{\rm #1}}}

\begin{document}

\title{Point-contact spectroscopy in neutron-irradiated Mg$^{11}$B$_2$}

\author{D. Daghero\email{E-mail:dario.daghero@infm.polito.it}}
\author{A.~Calzolari}
\author{G.A. Ummarino}
\author{M.~Tortello}
\author{R.S. Gonnelli}
\affiliation{Dipartimento di Fisica and CNISM, Politecnico di
Torino, 10129 Torino, Italy}
\author{V.A. Stepanov}
\affiliation{P.N. Lebedev Physical Institute, Russian Academy of
Sciences, 119991 Moscow, Russia}
\author{C. Tarantini}
\affiliation{CNR-INFM-LAMIA and Dipartimento di Fisica,
Universit\`{a} di Genova, 16146 Genova, Italy}
\author{P. Manfrinetti}
\affiliation{CNR-INFM-LAMIA and Dipartimento di Chimica e Chimica
Industriale,
Universit\`{a} di Genova, 16146 Genova, Italy }%
\author{E. Lehmann}
\affiliation{Paul Scherrer Institut, Dept. Spallation Neutron
Source SINQ , CH-5232 Villigen, Switzerland}%


\begin{abstract}
We report on recent results of point-contact spectroscopy
measurements in Mg$^{11}$B$_2$ polycrystals irradiated at different
neutron fluences up to $\Phi=1.4 \cdot 10^{20} \mathrm{cm}^{-2}$.
The point contacts were made by putting a small drop of Ag paint --
acting as the counterelectrode -- on the cleaved surface of the
samples. The gap amplitudes were extracted from the experimental
conductance curves, showing Andreev-reflection features, through a
two-band Blonder-Tinkham-Klapwijk fit and reported as a function of
the Andreev critical temperature of the junctions,
$T\ped{c}\apex{A}$. The resulting
$\Delta\ped{\sigma}(T\ped{c}\apex{A})$ and
$\Delta\ped{\pi}(T\ped{c}\apex{A})$ curves show a clear merging of
the gaps when $T\ped{c}\apex{A} \simeq 9$~K that perfectly confirms
the findings of specific-heat measurements in the same samples.
Anomalous contacts with $T\ped{c}\apex{A} > T\ped{c}$ (being
$T\ped{c}$ the bulk critical temperature) were often obtained,
particularly in samples irradiated at very high fluences. Their fit
gave a different dependence of $\Delta\ped{\pi}$ on
$T\ped{c}\apex{A}$. The possible origin of these contacts is
discussed in terms of local current-induced annealing and/or
nanoscale inhomogeneities observed by STM in the most irradiated
samples.
\end{abstract} \maketitle

\section{Introduction}\label{sect:intro}
Most of the present fundamental research on the two-band
superconductor MgB$_2$ is devoted to studying the effects of
substitutions and disorder on its properties. This interest in
exploring the ``neighborhood'' of the pure compound is justified in
part by the quest of a recipe for improving some of its properties
-- especially in view of power or electronic applications -- and in
part by the need of understanding at the best this unique example of
two-band phonon-mediated superconductor with a relatively high
$T_c$. As a matter of fact, the presence of two systems of bands
crossing the Fermi surface, each developing an energy gap below
$T\ped{c}$, has a number of intriguing consequences that make the
physics of MgB$_2$ unexpectedly rich and complex. One of these
aspects is the role of scattering by impurities. Due to the
different parity of the $\sigma$ and $\pi$ bands, scattering of
quasiparticles \emph{between} them is highly improbable in the pure
compound, while almost independent scattering rates exist
\emph{within} the bands. Intraband scattering has no effect on the
gaps or on $T\ped{c}$, but affects various properties of the
material, e.g. the critical field \cite{Gurevich} and the
magnetic-field dependence of the gaps \cite{Koshelev}. On this
basis, the ratio of the diffusivities in the two bands has been
experimentally evaluated in pure MgB$_2$
\cite{nostro_campo,Eskildsen,Bugoslavsky}.

According to an early prediction of the two-band model \cite{Liu},
the increase of interband scattering in a system like MgB$_2$ should
make the two gaps approach each other and finally merge into a
single BCS gap. However, observing this effect in a real material
has turned out to be more difficult than expected. The few chemical
substitutions that actually take place (e.g. C substitution for B,
Al or Mn substitution for Mg) give rise to lattice or electronic
effects that can mask the increase in disorder. For example, C and
Al substitutions also cause remarkable changes in the DOS at the
Fermi level, in the phonon frequencies, in the cell volume and so
on, with an obvious complication in the interpretation of the data.
In C-substituted single crystals the merging of the gaps has been
recently observed \cite{nostro_C} as arising from the interplay of
the band filling due to electron doping and an increasing amount of
interband scattering, probably due to extrinsic reasons
\cite{Kortus,John_C}.

The controlled damaging of the compound by means of irradiation
allows partly overcoming these difficulties, even though the
irradiated material is far from being an ``ideal'' disordered
version of MgB$_2$. In particular, side effects of irradiation are
the creation of Li atoms and He nuclei in the lattice (due to the
thermal neutron capture by $^{10}$B nuclei), the reduction in the
partial DOS of the $2p_{x,y}$ states \cite{Geraschenko}, the
anisotropic expansion of the crystal lattice \cite{Putti_APL}. The
effect of neutron irradiation (up to very high fluences) on the
energy gaps of MgB$_2$ has been recently studied by means of
specific-heat measurements \cite{Putti_gap}, showing the achievement
of single-gap superconductivity in samples with $T\ped{c}$ as low as
11~K.

In this paper, we present and discuss the results of point-contact
spectroscopy measurements (in the Andreev-reflection regime) in the
same neutron-irradiated Mg$^{11}$B$_2$ samples studied in Ref.
\onlinecite{Putti_gap}. We will show that the gap amplitudes
measured by PCS agree very well with those given by specific-heat
measurements and we will discuss the experimental trend of the gaps
within the two-band Eliashberg theory.

Moreover, we will report on the anomalous features of a large
number of contacts whose Andreev critical temperature
$T\ped{c}^{A}$ is greater than the bulk $T\ped{c}$. These contacts
feature very good Andreev-reflection conductance curves that were
very well fitted by the two-band Blonder-Tinkham-Klapwijk (BTK)
model to extract the gap amplitudes $\Delta\ped{\sigma}$ and
$\Delta\ped{\pi}$. Once reported as a function of
$T\ped{c}\apex{A}$, $\Delta\ped{\pi}$ has a completely different
trend with respect to that reported in Ref. \onlinecite{Putti_gap}
and observed by PCS in ``standard'' contacts. We will discuss this
odd result in terms of local nanoscale inhomogeneities of the
material and/or local annealing due to the technique we used to
tune the properties of our ``soft'' point-contact junctions.

\section{Experimental details}\label{sect:exp}
The procedure for sample fabrication and irradiation is reported in
detail elsewhere \cite{Putti_gap,Putti_APL}. The samples were
prepared by direct synthesis from pure elements, using in particular
isotopically-enriched $^{11}$B (99.95\% purity) with a residual
$^{10}$B concentration lower than 0.5\% so as to make the
penetration depth of thermal neutrons greater than the sample
thickness \cite{Putti_gap}. The samples we measured had been
irradiated at the Paul Scherrer Institute (PSI) in Villigen,
Switzerland. For simplicity and ease of comparison, let us label
them as in Ref.\onlinecite{Putti_gap}, i.e. P0 (pristine
Mg$^{11}$B$_2$), P3 (fluence $\Phi=7.6 \cdot 10^{17}
\mathrm{cm}^{-2}$), P3.7 ($\Phi=5.5 \cdot 10^{18}
\mathrm{cm}^{-2}$), P4 ($\Phi=1.0 \cdot 10^{19} \mathrm{cm}^{-2}$)
and P6 ($\Phi=1.4 \cdot 10^{20} \mathrm{cm}^{-2}$). Many of their
structural and transport properties are reported in Refs.
\onlinecite{Putti_APL} and \onlinecite{Putti_gap}. The bulk critical
temperatures, defined as  $T\ped{c}\equiv T\ped{90\%}$ of the
superconducting transition measured by susceptibility, are: $38.8$~K
for P0, $35.6$~K for P3, $25.8$~K for P3.7, $20.7 $~K for P4,
$8.7$~K for P6. The width of the superconducting transition, defined
as $\Delta T\ped{c}(10\%-90\%)$, varies from 0.3 K (in P0 and P3) up
to a maximum of 0.9 K (in sample P4)\cite{Putti_gap}. The transition
remains rather sharp also in the most irradiated sample, which
indicates a highly homogeneous defect distribution even at the
highest fluence. This homogeneity is also confirmed by the sharp
X-ray diffraction peaks for the (002) and (110) reflections reported
in Ref. \onlinecite{Tarantini}, which also indicate an anisotropic
expansion of the cell parameters, more pronounced along the
\textit{c} axis (up to 1\%). The residual resistivity increases by
two orders of magnitude (from 1.6 $\mu\Omega\cdot$cm for P0 to 130
$\mu\Omega\cdot$cm for P6) with a corresponding reduction in the
residual resistivity ratio (RRR).

The point contacts were made by placing a small drop of silver paint
on the freshly cleaved surface of the samples \cite{nostroPRL}. The
conductance curves, d$I$/d$V$ vs. $V$, were obtained by numerical
differentiation of the measured $I-V$ curve. In all cases, we
studied the temperature dependence of the curves, which show clear
Andreev-reflection features, so as to determine the critical
temperature of the junction (in the following referred to as the
``Andreev critical temperature'', $T\ped{c}\apex{A}$). Strictly
speaking, in fact, $T\ped{c}\apex{A}$ rather than the bulk
$T\ped{c}$ is the critical temperature to be related to the local
gap amplitudes measured in a given contact. In point-contact
spectroscopy, $T\ped{c}\apex{A}$ can correspond to any temperature
between the onset and the completion of the superconducting magnetic
transition. Therefore, one usually has $T\ped{c}\apex{A} = T\ped{c}$
within the experimental broadening of the superconducting
transition, i.e. $T\ped{0\%} \leq T\ped{c}\apex{A} \leq
T\ped{100\%}$ (let us recall that here we defined $T\ped{c}\equiv
T\ped{90\%}$). In irradiated samples, this actually occurs in a
subset of contacts we will call ``standard'' contacts. The
conductance curves were divided by the normal-state conductance and
then fitted with a two-band BTK model in which the conductance
through the junction is expressed by $G = (1-\omega _\pi)G_\sigma +
\omega _\pi G_\pi$, $G_\sigma$ and $G_\pi$ being the partial
$\sigma$- and $\pi$-band conductances, and $\omega _\pi$ the weight
of the $\pi$-band contribution \cite{BTK,nostroPRL}. The model
contains as adjustable parameters the gap amplitudes
$\Delta\ped{\sigma}$ and $\Delta\ped{\pi}$, the barrier parameters
$Z\ped{\sigma}$ and $Z\ped{\pi}$, the phenomenological broadening
parameters $\Gamma\ped{\sigma}$ and $\Gamma\ped{\pi}$, plus the
weight $w\ped{\pi}$. The broadening parameters enter in the
definition of the density of states in the usual way, i.e.
\[N(E)= \Re \left(\frac
{E-i\Gamma}{\sqrt{(E-i\Gamma)^2-\Delta^2}}\right).\]
In this context, they account for both intrinsic (i.e. finite
lifetime of quasiparticles) and extrinsic (related to the
technique and the nature of the contacts) phenomena that smear out
the conductance curves. $Z\ped{\sigma,\pi}$ are related to the
potential barrier height at the interface and to the mismatch in
the Fermi velocity $v\ped{F}$ between the two sides of the
contact. Owing to the different values of $v\ped{F}$ in the
$\sigma$ and $\pi$ bands, we allow $Z\ped{\sigma} \neq
Z\ped{\pi}$. Finally, $w\ped{\pi}$ is predicted to range from 0.66
to 0.99 for perfectly directional tunneling in pure MgB$_2$
\cite{Brinkman}, depending on the angle of current injection with
respect to the $ab$ planes. In the absence of specific predictions
in samples with reduced anisotropy, we kept $w\ped{\pi}$ in the
same range as in Refs. \onlinecite{nostroPRL,nostro_campo}.

We generally selected contacts with rather high values of the
normal-state resistance $R\ped{N}$, corresponding to small values of
the contact size $a$ which has to be smaller than the electronic
mean free path $\ell$ for energy-resolved spectroscopy to be
possible. The limit condition $a \ll \ell$ defines the so-called
ballistic regime of conduction \cite{Duif}. Based on Ref.
\onlinecite{dips}, we also required the conductance curves of our
point contacts not to present dips, which are the hallmark of a
breakdown of the conditions for ballistic conduction at finite
voltage and signal the presence of heating in the contact region.
When the contact resistance was too small, or its conductance did
not show clear Andreev-reflection features, we were able to change
the contact characteristics (in a surprisingly repeatable way) by
applying short voltage or current pulses to the junction itself. In
some cases, we also used the magnetic field to clarify whether one
or two gaps were present, as explained in detail elsewhere
\cite{nostroPRL} and in the following.

\section{Results in standard contacts}\label{sect:standard}
\begin{figure}[t]
\begin{center}
\includegraphics[keepaspectratio, width=0.8\columnwidth]{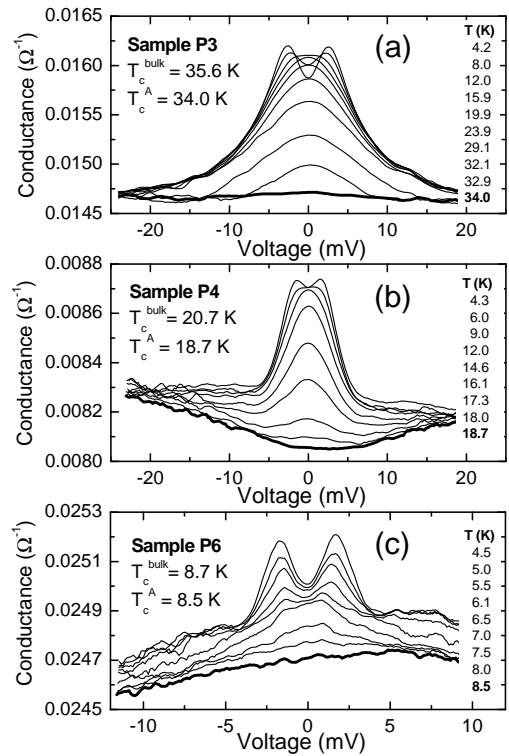}
\end{center}
\vspace{-5mm} \caption{\small{Temperature dependence of the raw
conductance curves measured in three contacts on samples P3, P4
and P6, whose bulk critical temperature is indicated. Thicker
lines indicate the normal-state conductance, which, in most of the
contacts, is practically temperature-independent and is reached
when $T = T\ped{c}^{A}$. The temperature of each curve is also
indicated in the legend.}} \label{Fig:raw_standard}
\end{figure}

Fig.~\ref{Fig:raw_standard} reports an example of the raw
conductance curves measured as a function of the temperature in
three point contacts made on samples P3, P4 and P6. The
low-temperature curves clearly show the typical Andreev-reflection
features -- in particular, the two symmetric maxima at $\pm
V\ped{peak}$ approximately corresponding to the edges of the small
gap. The large gap-features are usually less clear even in pure
MgB$_2$ \cite{nostroPRL} and in disordered samples they are
difficult to see. The thick curve in each panel indicates the
normal-state conductance and the relevant temperature is thus chosen
as the $T\ped{c}^{A}$ of the contacts. Note that, in all cases,
$T\ped{c}^{A}$ lies between the begin and the end of the magnetic
superconducting transition.

In Fig.~\ref{Fig:raw_standard}(a) and (b) the shape of the
normal-state conductance curves  and the absence of dips \cite{dips}
in the superconducting state indicate that no heating occurs in the
junctions, which are thus in the ballistic regime (i.e., the contact
size $a$ is smaller than the mean free path $\ell$) and
energy-resolved spectroscopy is possible. Here and in the following
$\ell$ is defined through the relation $\ell^{-1}= \ell\ped{e}^{-1}+
\ell\ped{i}^{-1}$, where $\ell\ped{e}$ and $\ell\ped{i}$ are the
elastic and inelastic mean free paths, respectively (at sufficiently
low temperature, $\ell \simeq \ell\ped{e}$).

In Fig.~\ref{Fig:raw_standard}(c) small dips at $|V|>|V\ped{peak}|$
suggest that the contact might actually be in the diffusive regime
(i.e. $\ell\ped{e}< a < \Lambda$, being $\Lambda$ the diffusion
length \cite{Duif}) in which energy-resolved spectroscopy is still
possible. In this regime, the resistance of a point contact between
two normal metals -- in the hypothesis that the Fermi velocity of
the metals are almost equal and the resistivity of one metal (here
Ag) is much smaller than the resistivity of the other (here sample
P6) -- can be expressed by \cite{Wexler}
\begin{equation}
R\ped{N}= \frac{4 \rho \ell}{3 \pi a^2}+ \Gamma(k) \frac{\rho}{4a}
\label{eq:Wexler}
\end{equation}
where the first term is the Sharvin resistance for ballistic
conduction and the second one is the Maxwell resistance for the
thermal regime multiplied by a function of the Knudsen ratio
$k=\ell/a$. This function, $\Gamma$, is always of the order of
unity. $\rho$ is the normal-state resistivity of the irradiated
sample (that we will identify with the residual resistivity) and
$\ell$ is the mean free path (evaluated in
Ref.\onlinecite{Tarantini}). In the second term, the contribution of
the first half of the contact (the Ag counterelectrode) has been
neglected \cite{Naidyuk,Gloos} due to the much smaller resistivity
of Ag with respect to that of sample P6. At $T < T\ped{c}$ the
irradiated MgB$_2$ is superconducting and, thus, the contribution of
the second term in eq.~\ref{eq:Wexler} should disappear.
Nevertheless, it is easy to show that even at very low bias voltages
-- of the order of the energy gap, here about 1 meV as we will
discuss later -- the current density in the contact is higher than
the critical current density and thus tends to drive normal a small
volume of the superconductor pushing back the NS boundary a short
distance \cite{Waldram}. If the size of this normal region is
smaller than the coherence length the spectroscopy of the gap is
still possible \cite{Deutscher}, but the second term in
eq.~\ref{eq:Wexler} starts playing a role and a small,
voltage-dependent and temperature-dependent heating appears in the
contact. The small vertical shift in the conductance curves shown in
Fig.~\ref{Fig:raw_standard}(c) on increasing temperature confirms
this picture indicating that the temperature-dependent resistivity
of the material plays a role in the contact resistance. Heating of
the contact region generally gives rise to an apparent decrease of
the critical temperature of the contact $T\ped{c}\apex{A}$ with
respect to the bulk $T\ped{c}$. However, here the superconducting
features disappear at some temperature between 8.0 and 8.5 K, which
is only slightly smaller than $T\ped{c}=8.7$ K. Hence, we can
conclude that a very moderate heating is likely to occur in the
contact shown in Fig.~\ref{Fig:raw_standard}(c) and it can be safely
neglected as long as the voltage drop across the junction is of the
order of $V\ped{peak}$. We will show in greater detail in a
following section that the two conditions described above
(normal-region size $< \xi$ and very small heating) are compatible
with the curves shown in Fig.~\ref{Fig:raw_standard}(c) only if
parallel diffusive nanocontacts are supposed to be present in the
contact region.

\begin{figure}[t]
\begin{center}
\includegraphics[keepaspectratio, width=0.8\columnwidth]{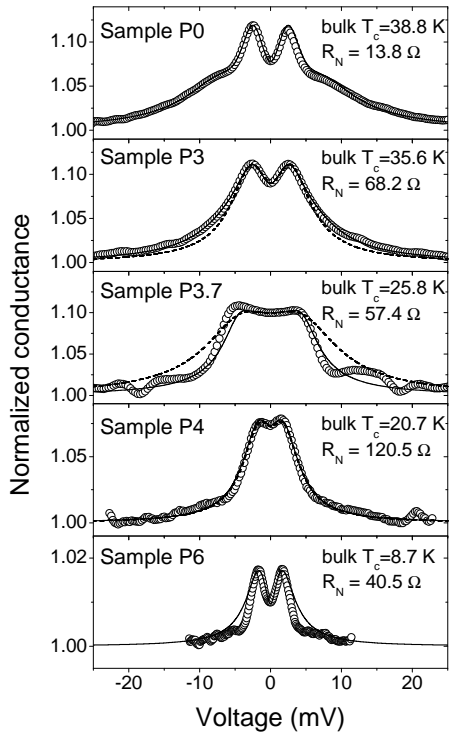}
\end{center}
\vspace{-5mm} \caption{\small{Normalized conductance curves
(symbols) of different point contacts on neutron-irradiated
Mg$^{11}$B$_2$ polycrystals at T=4.2 K. The curves are labeled with
the name of the samples and the relevant bulk $T\ped{c}$ according
to Ref.~\onlinecite{Putti_gap}. The values of the normal-state
resistance are also indicated. Solid lines are the best-fitting
curves given by the appropriate BTK model (two-band in samples P0 to
P4, single-band in sample P6). Dashed lines shown for samples P3,
P3.7 and P4 represent the single-band best-fitting curve to be
compared to the two-band fit. In the case of sample P4, solid and
dashed lines are almost superimposed. }} \label{Fig:totale}
\end{figure}

Fig.~\ref{Fig:totale} reports an example of experimental, normalized
conductance curve (symbols) for each sample. Notice that the
horizontal scale is the same for all the panels, so as to highlight
the shrinking of the Andreev-reflection structures on increasing the
neutron fluence -- which indicates, in turn, a decrease in the
amplitude of the gaps. While in the top curve (sample P0) the
presence of peaks and shoulders clearly witnesses the existence of
two gaps, in the irradiated samples this evidence is lacking, due to
a progressive broadening of the curves accompanied by a reduction in
their height. The same happens in doped MgB$_2$ \cite{nostro_C,
nostro_Al}. In all these cases, the existence of two gaps can be
inferred from the fit of the curves with the BTK model or evidenced
by the application of a magnetic field \cite{nostro_campo}.

The BTK curves that best fit the experimental data are shown in
Fig.~\ref{Fig:totale} as solid lines. In sample P0 (pristine
Mg$^{11}$B$_{2}$), the fit can only be obtained with the two-band
BTK model. In samples P3 and P3.7, the two-band fit works better
than the single-band one, since it reproduces both the width of the
Andreev-reflection structures and the position of the peaks, while
the single-band fit (dashed lines) does not. The curve measured in
sample P4 admits both a single-band and a two-band BTK fit, which
are almost equally good -- as a matter of fact, dashed and solid
lines are almost superimposed in this case. In sample P6, the dips
at $|V|> |V\ped{peak}|$ modify the shape of the curve so that asking
the model to fit the curve in this region is nonsense. In these
conditions, the parameters of the BTK model should be adjusted so as
to fit the conductance maxima and the zero-bias dip between them.
The single-band BTK model is sufficient to accomplish this task very
well (see the line in the bottom panel of Fig.~\ref{Fig:totale}). If
a two-gap fit is tried, the values of $\Delta\ped{\pi}$ and
$\Delta\ped{\sigma}$ turn out to be so close to each other to be
practically identical. Hence, in this sample the existence of a
single gap can be safely concluded, in agreement with the findings
of Ref. \onlinecite{Putti_gap}.

\begin{figure}[t]
\begin{center}
\includegraphics[keepaspectratio, width=0.9\columnwidth]{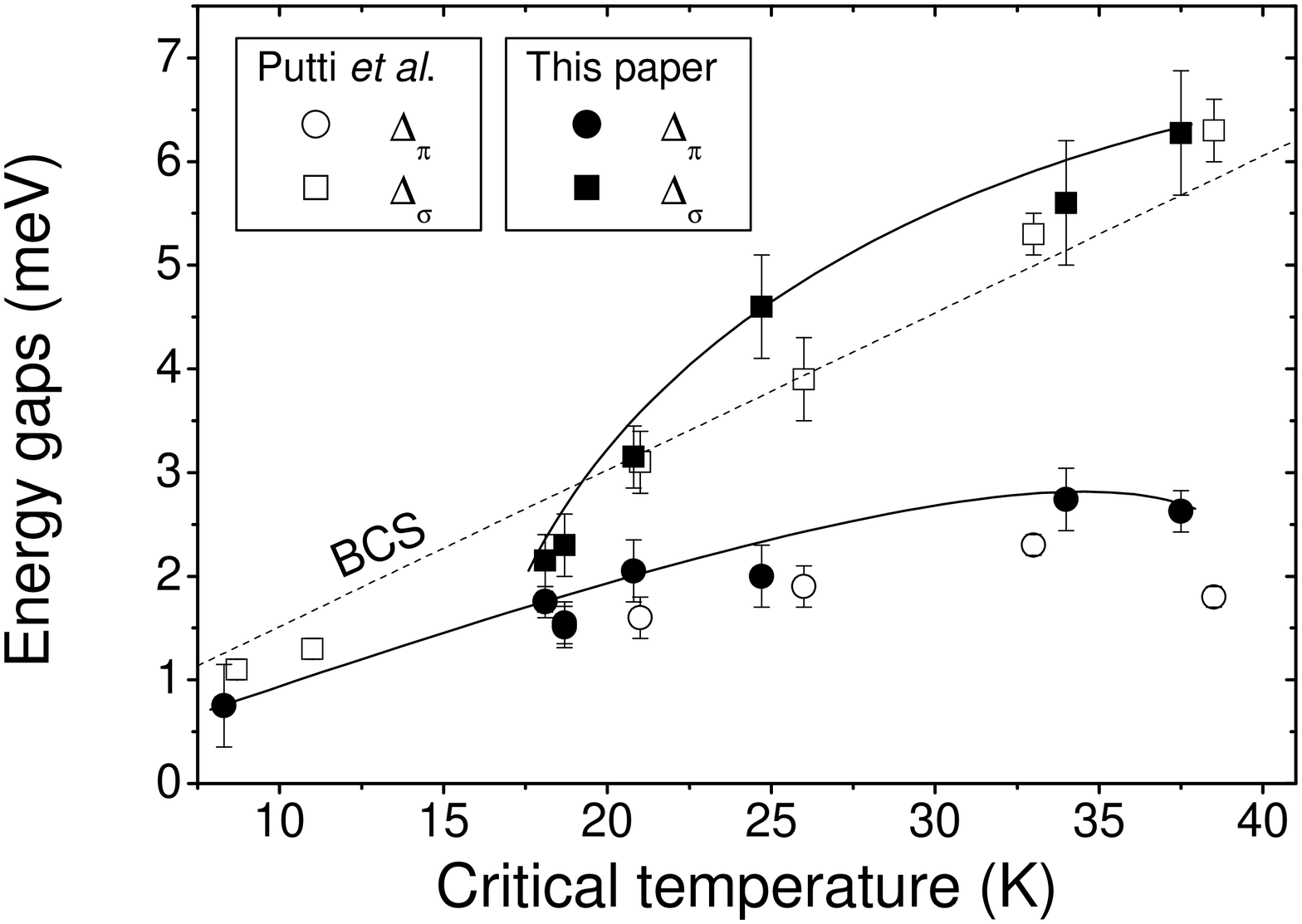}
\end{center}
\vspace{-5mm} \caption{\small{Black symbols: gap amplitudes from PCS
measurements, as a function of the Andreev critical temperature,
$T\ped{c}\apex{A}$. Open symbols: gap amplitudes from specific-heat
measurements \cite{Putti_gap} as a function of the critical
temperature measured by specific heat. Solid lines are only guides
to the eye, while the straight dashed line indicates the value of
the gap for a conventional, BCS superconductor.}}
\label{fig:gaps_vs_Tc}
\end{figure}

The gap amplitudes extracted from the fit of the curves shown in
Fig.~\ref{Fig:totale} (and of other curves not reported here,
measured in different contacts on the same samples) are plotted in
Fig.~\ref{fig:gaps_vs_Tc} as a function of $T\ped{c}^{A}$ (black
symbols). In the region around 18-19 K, the gap amplitudes resulting
from the \emph{two-band} fit of the conductance curves are shown,
but it should be borne in mind that a single-gap fit is possible as
well in this region, giving a gap value $\Delta \simeq
\Delta\ped{\pi}$. In the same figure, the gap amplitudes extracted
from the fit of specific-heat measurements \cite{Putti_gap} are also
shown (open symbols). The agreement between the two sets of data is
good in the whole range of critical temperatures, especially if one
takes into account that: i) PCS is a local, surface-sensitive
technique while specific heat is a bulk property; ii) the gap values
obtained by PCS are correctly plotted versus the Andreev critical
temperature of the contacts, $T\ped{c}^{A}$, while those taken from
Ref.~\onlinecite{Putti_gap} are reported as a function of the
specific-heat $T\ped{c}$.

The trend shown in Fig.~\ref{fig:gaps_vs_Tc} clearly indicates a
transition from two-band to single-band superconductivity at high
neutron fluences. That the heavily-irradiated material undergoes
deep changes above a certain neutron fluence ($\Phi \simeq 10^{19}$
cm$^{-2}$) is confirmed by the steep decrease in $T\ped{c}$ and
$B\ped{c2}$, by the increase in the cell parameters $a$ and $c$
\cite{Putti_APL,Tarantini} and by the observed decrease in the
$\sigma$-band DOS \cite{Geraschenko}. Incidentally, it is
interesting to note the initial, small increase in
$\Delta\ped{\pi}$, which is the hallmark of an increase in the
scattering between bands.

We tried to reproduce the experimental trend of the gaps
$\Delta\ped{\pi}$ and $\Delta\ped{\sigma}$ reported in
Fig.~\ref{fig:gaps_vs_Tc} within the two-band Eliashberg theory. A
complete fit of the $\Delta\ped{\pi}(T\ped{c}\apex{A})$ and
$\Delta\ped{\sigma}(T\ped{c}\apex{A})$ curves is actually impossible
since in a certain range of critical temperatures both gaps are
smaller than the BCS value (see Fig.~\ref{fig:gaps_vs_Tc}) until a
BCS-like gap ratio is almost recovered at $T\ped{c}\simeq 9$~K, when
$\Delta \simeq 1$~meV. An energy gap smaller than the BCS value has
indeed been observed in disordered, conventional superconductors
\cite{Vaglio} and the same might occur in a two-band system, but it
is strictly forbidden within the Eliashberg theory and no
explanation for these findings has been given yet. Once established
this point, one can proceed with the fit.

\begin{figure}[t]
\begin{center}
\includegraphics[keepaspectratio, width=\columnwidth]{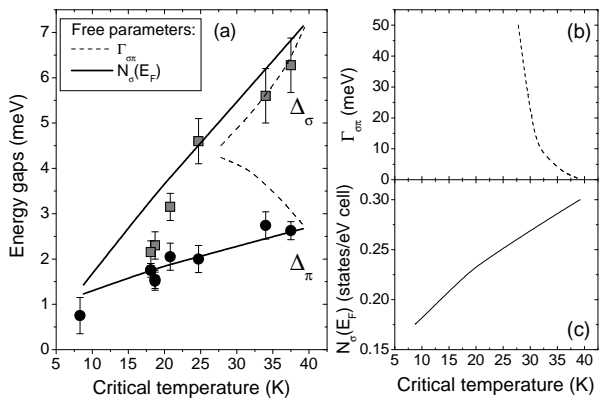}
\end{center}
\vspace{-5mm} \caption{(a) Symbols: gap amplitudes from PCS
measurements, as a function of the Andreev critical temperature,
$T\ped{c}\apex{A}$. Lines: gap amplitudes vs. $T\ped{c}\apex{A}$
calculated by solving the two-band Eliashberg equations using as
adjustable parameters: i) the interband scattering
$\Gamma\ped{\sigma \pi}$ alone (dashed lines); ii) the $\sigma$-band
DOS $N\ped{\sigma}(E\ped{F})$ alone (thin solid lines). (b) The
values of $\Gamma\ped{\sigma \pi}$ that give rise to the dashed
curve of panel (a) (previous case (i)). (c) The
$N\ped{\sigma}(E\ped{F})$ vs. $T\ped{c}$ curve necessary to fit the
data (previous case (ii)).} \label{fig:Eliashberg_standard}
\end{figure}

The simplest approach is to consider the irradiated material as if
it was only ``disordered MgB$_2$'', thus neglecting the changes in
the DOS, in the phonon frequencies and in the cell volume, and only
increasing the interband scattering $\Gamma\ped{\sigma \pi}$. Once
the value of this single parameter is chosen to reproduce the
critical temperature of a given sample, no further degrees of
freedom are left to reproduce the gap amplitudes. The resulting
curves are shown in Fig.~\ref{fig:Eliashberg_standard}(a) as dashed
lines and the corresponding values of $\Gamma\ped{\sigma \pi}$ are
reported in Fig.~\ref{fig:Eliashberg_standard}(b). Notice that very
high values of $\Gamma\ped{\sigma \pi}$ would be necessary to
suppress $T\ped{c}$ below 30 K and make the gaps merge. Such an
intense interband scattering is never observed in real systems and
is probably not physical. In doped MgB$_2$, for example, the
suppression of $T\ped{c}$ and $\Delta\ped{\sigma}$ is mainly due to
other effects (typically a reduction in the DOS\cite{John_C}) and
even if $\Gamma\ped{\sigma \pi}$ moderately increases (usually
remaining smaller than 10 meV) its effects are partially masked, so
that a tendency of $\Delta\ped{\pi}$ to remain constant or slightly
increase is at most observed \cite{nostro_C,John_C}. The present
case is not very different since, as previously pointed out, neutron
irradiation has ``side effects'' such as sizeable changes in the
$\sigma$ DOS and in the cell parameters that are not included in
this description.

The opposite approach for the fit of the experimental gap values of
Fig.\ref{fig:Eliashberg_standard}(a) thus consists in disregarding
the effect of disorder (scattering) and only taking into account the
change in the $\sigma$ DOS at the Fermi level,
$N\ped{\sigma}(E\ped{F})$ \cite{Geraschenko}. A reasonable fit of
the experimental $\Delta\ped{\sigma}(T\ped{c}\apex{A})$ and
$\Delta\ped{\pi}(T\ped{c}\apex{A})$ curves (with the general
theoretical limitation that the gap ratios cannot be both smaller
than the BCS one) is indeed obtained in this way, as indicated by
the solid lines in Fig.~\ref{fig:Eliashberg_standard}(a). That using
a single parameter one can reproduce in such a good way the values
of the two gaps \emph{and} the critical temperature is, by itself, a
good result and indicates that $N\ped{\sigma}(E\ped{F})$ is largely
dominant in determining the observed gap trend.  This conclusion is
consistent with the findings of Ref. \onlinecite{Geraschenko} where
the depression of $T\ped{c}$ down to about 10 K was justified by
inserting in the McMillan formula the reduced DOS (about 25\% of the
value in pristine MgB$_2$) measured by NMR.
Fig.~\ref{fig:Eliashberg_standard}(c) reports the $T\ped{c}\apex{A}$
dependence of the $\sigma$-band DOS necessary to fit our PCS data
(solid line).

To account for the initial increase in $\Delta\ped{\pi}$ (also
clearly shown by specific-heat measurements\cite{Putti_gap}) a small
amount of interband scattering must be inserted in the model. In
particular, in our case $\Gamma\ped{\sigma \pi}$ should be about 0.7
meV when $T\ped{c}$=35 K and should saturate at a constant value (no
more than 2 meV) at low $T\ped{c}$. It is clear, however, that even
including in the model all the possible effects of neutron
irradiation, the agreement with the data can be hardly improved due
to the aforementioned anomaly of the gap values that are both
smaller than the BCS value in the $T\ped{c}\apex{A}$ range between
$\simeq 9$~K and $\simeq 20$~K.

\section{Results in anomalous contacts}\label{sect:anomalous}
\begin{figure}[t]
\begin{center}
\includegraphics[keepaspectratio, width=0.8\columnwidth]{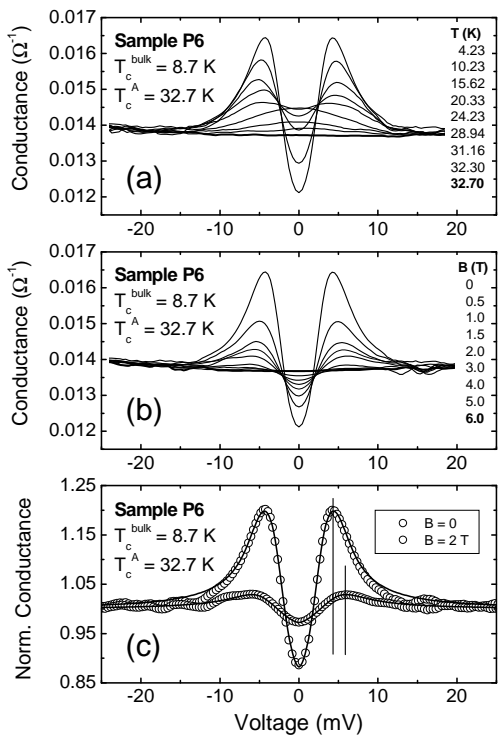}
\end{center}
\vspace{-5mm} \caption{\small{(a) Temperature dependence of the raw
conductance curves measured in a contact with $T\ped{c}^{A}$=32.7 K
obtained on sample P6 (bulk $T\ped{c}$=8.7 K). The thick line is the
normal-state conductance. (b) Magnetic-field dependence of the same
conductance curve as in (a). Again, the thick line is the
normal-state conductance. (c) Comparison of the conductance curves
in zero field and in a magnetic field of 2 T. The outward
displacement of the conductance peaks is highlighted by vertical
lines.}} \label{Fig:raw_anomali}
\end{figure}

The percentage of ``standard'' contacts (as defined in the previous
section) is equal to 100\% in samples P0 and P3, but fast decreases
in more irradiated samples. For example, it is about 70\% in sample
P3.7, 30\% in sample P4 and becomes as small as 10\% in sample P6.
The remaining contacts are ``anomalous'' in the sense that their
$T\ped{c}^{A}$ exceeds $T\ped{c}$, which is clearly related to some
kind of intrinsic or induced inhomogeneity in the samples.
Fig.~\ref{Fig:raw_anomali} reports the conductance curves of one of
such anomalous contacts on the most irradiated sample (P6, bulk
$T\ped{c}=8.7$~K). The normal-state resistance of the contact was
$R\ped{N} = 71\, \Omega$. The temperature dependence of its
conductance curve, reported in Fig.~\ref{Fig:raw_anomali}(a),
clearly shows that the Andreev-reflection features persist well
above the bulk critical temperature and disappear at
$T\ped{c}^{A}=32.7$~K, which is more than three times the bulk
$T\ped{c}$ measured by susceptibility. The thick line in
Fig.~\ref{Fig:raw_anomali}(a) indicates the normal-state conductance
curve. A fit of the low-temperature curve with the BTK model
unambiguously shows the presence of two gaps whose values are
similar (but not identical) to those obtained in standard contacts
with the same $T\ped{c}^{A}$.

To further enlighten this point, we applied to the junction a
magnetic field perpendicular to the direction of main current
injection, and studied the behavior of the conductance curves on
increasing the field intensity. In pure MgB$_2$, using a magnetic
field allowed us to separate the partial contributions of the
$\sigma$ and $\pi$ bands to the conductance across the junction
\cite{nostroPRL,nostro_campo}. In doped MgB$_2$, the complete
separation is not always possible but, if two gaps are present, an
outward shift of the conductance maxima occurs when the smaller gap
is strongly suppressed by the field \cite{nostro_C,nostro_Mn}.
Fig.~\ref{Fig:raw_anomali}(b) reports the raw conductance curves of
the same contact as in Fig.~\ref{Fig:raw_anomali}(a), measured in a
magnetic field of increasing intensity. Again, the thick line
corresponds to the normal-state conductance curve obtained here at
$B =$ 6 T (notice that it is identical to that of panel (a)). It is
clearly seen that, at lower fields, the conductance peaks shift
towards higher energies -- a behavior that cannot be explained
within a single-band model and arises from the stronger suppression
of the $\pi$-band gap by the magnetic field
\cite{Koshelev,nostro_campo}. The curves measured with $B=0$ and
$B=2$~T are reported, after normalization, in
Fig.~\ref{Fig:raw_anomali}(c) together with the relevant two-band
BTK fit. The shift of the conductance peaks is indicated by the two
vertical lines. The values of the best-fitting parameters are the
following: $\Delta\ped{\pi}=3.38$~meV, $\Gamma\ped{\pi}=1.35$~meV,
$Z\ped{\pi}=0.74$, $\Delta\ped{\sigma}=5.00$~meV,
$\Gamma\ped{\sigma}=1.20$~meV, $Z\ped{\sigma}=0.9$ for the
zero-field curve; $\Delta\ped{\pi}=0.6$~meV,
$\Gamma\ped{\pi}=2.05$~meV, $Z\ped{\pi}=0.74$,
$\Delta\ped{\sigma}=4.85$~meV, $\Gamma\ped{\sigma}=2.75$~meV,
$Z\ped{\sigma}=0.9$ for the curve in magnetic field. The weight
$w\ped{\pi}$ was taken equal to 0.8, as usual in polycrystalline
samples.

\begin{figure}[t]
\begin{center}
\includegraphics[keepaspectratio, width=0.8\columnwidth]{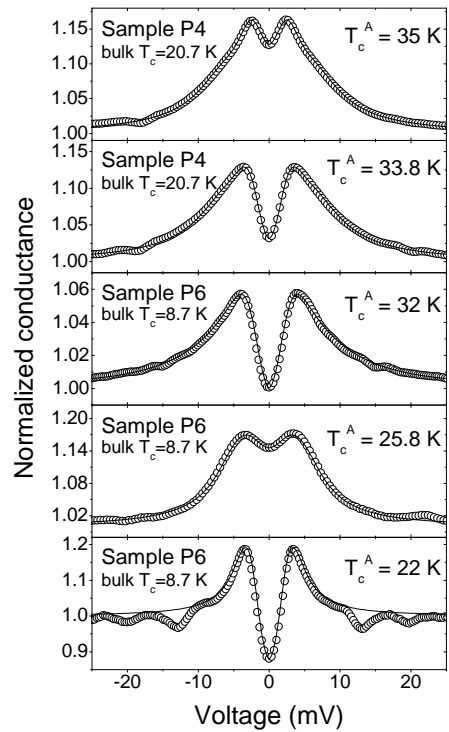}
\end{center}
\vspace{-5mm} \caption{\small{Normalized conductance curves
(symbols) of different anomalous point contacts on
neutron-irradiated Mg$^{11}$B$_2$ polycrystals at T=4.2 K. The
curves are labeled with the Andreev critical temperature
$T\ped{c}^{A}$ but the name of the samples and the relevant bulk
$T\ped{c}$ are also indicated. Lines are the best-fit curves given
by the two-band BTK model.}} \label{Fig:totale_anomali}
\end{figure}

\begin{figure}[t]
\begin{center}
\includegraphics[keepaspectratio, width=0.9\columnwidth]{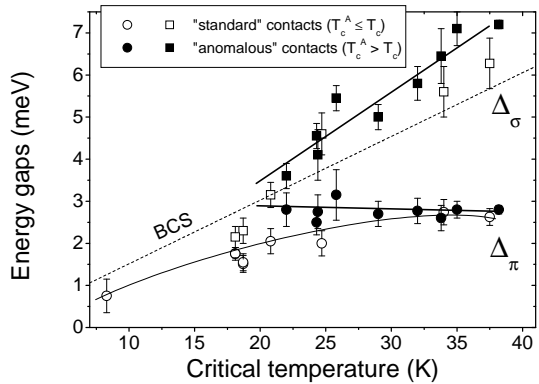}
\end{center}
\vspace{-5mm} \caption{\small{Gap amplitudes from PCS measurements,
as a function of the Andreev critical temperature,
$T\ped{c}\apex{A}$, in anomalous contacts ($T\ped{c}\apex{A}>
T\ped{c}$) (black symbols) compared to the gaps in standard contacts
(open symbols). Solid lines are only guides to the eye, while the
straight dashed line indicates the value of the gap for a
conventional, BCS superconductor.}} \label{fig:gaps_vs_Tc_anomali}
\end{figure}

Fig.~\ref{Fig:raw_anomali} clearly shows that, apart from the high
value of $T\ped{c}^{A}$, the anomalous contacts present very regular
conductance curves, with a smooth dependence on magnetic field and
temperature. Other examples of normalized conductance curves of
anomalous contacts with $T\ped{c}^{A}$ ranging from 35 K down to 22
K are reported in Fig.~\ref{Fig:totale_anomali}, together with the
relevant two-band BTK fits. The agreement between experimental data
and fitting curves is remarkably good. The gap amplitudes extracted
from these fits (and from the fit of the curves in other anomalous
contacts) are reported as a function of the local critical
temperature $T\ped{c}^{A}$ in Fig.~\ref{fig:gaps_vs_Tc_anomali}
(black symbols). The behavior of the gaps measured in standard
contacts is reported for comparison (open symbols). It is clear
that, while the $\Delta\ped{\sigma}$ vs. $T\ped{c}^{A}$ curve is
rather similar in standard and anomalous contacts, the trend of
$\Delta\ped{\pi}$ is fairly different. In anomalous contacts, the
small gap $\Delta\ped{\pi}$ tends to remain constant or slightly
increases on decreasing $T\ped{c}^{A}$, which generally indicates a
substantial increase in the interband scattering
\cite{Kortus,John_C}. Extrapolating the experimental curves to lower
critical temperatures suggests that the two gaps might tend to a
common value of about 3 meV and reach it at a critical temperature
of about 18 K, as in C-doped MgB$_2$ single crystals
\cite{nostro_C}. Finally, it is worthwhile to note that, in
anomalous contacts, the BCS rule for the gap ratio is no longer
violated.

\begin{figure}[t]
\begin{center}
\includegraphics[keepaspectratio, width=\columnwidth]{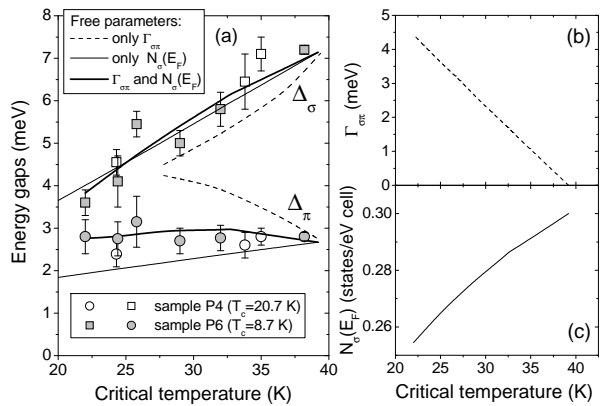}
\end{center}
\vspace{-5mm} \caption{(a) Gap amplitudes from PCS measurements in
anomalous contacts (symbols), as a function of the Andreev critical
temperature $T\ped{c}\apex{A}$. Open and filled symbols refer to
sample P4 and P6, respectively. Lines: theoretical curves obtained
within the Eliashberg theory by using as adjustable parameter only
$\Gamma\ped{\sigma \pi}$ (dashed lines), only
$N\ped{\sigma}(E\ped{F})$ (thin solid lines) or both of them (thick
solid lines). The values of $\Gamma\ped{\sigma \pi}$ and
$N\ped{\sigma}(E\ped{F})$ that give the thick solid lines in panel
(a) are reported as a  function of the critical temperature in
panels (b) and (c), respectively. } \label{fig:Eliashberg_anomali}
\end{figure}

The experimental trend of the gaps in anomalous contacts as a
function of $T\ped{c}^{A}$ can be analyzed within the two-band
Eliashberg theory, as we did for standard contacts. Again, the first
and simplest possibility consists in keeping all the parameters
fixed to their values in pristine MgB$_2$ and simply increasing the
interband scattering rate $\Gamma\ped{\sigma \pi}$ to simulate the
disorder due to irradiation. The theoretical curves are reported in
Fig.~\ref{fig:Eliashberg_anomali}(a) as dashed lines, and the
corresponding values of $\Gamma\ped{\sigma \pi}$ are the same we
already showed in Fig.~\ref{fig:Eliashberg_standard}(b). As in
standard contacts, the experimental values of $\Delta\ped{\pi}$ are
incompatible with this simple picture. As we did for standard
contacts, the next step towards a theoretical reproduction of the
experimental data is to allow variations in the $\sigma$-band
density of states at the Fermi level, neglecting the increase in
interband scattering. The best fit of the experimental gaps versus
$T\ped{c}$ is obtained by decreasing almost linearly
$N\ped{\sigma}(E\ped{F})$ from 0.30 down to 0.23 states/(eV unit
cell) while $T\ped{c}$ ranges from 38.8 K to 20 K. The resulting
curves, reported as thin solid lines in panel (a), clearly do not
follow the experimental values of $\Delta\ped{\pi}$. A much better
result can be obtained by using both $N\ped{\sigma}(E\ped{F})$ and
$\Gamma\ped{\sigma \pi}$ as adjustable parameters to fit the
experimental data. The best-fitting
$\Delta\ped{\sigma}(T\ped{c}^{A})$ and
$\Delta\ped{\pi}(T\ped{c}^{A})$ curves are reported in
Fig.~\ref{fig:Eliashberg_anomali}(a) (thick solid lines) while the
values of the relevant parameters are reported as a function of
$T\ped{c}^{A}$ in Fig.~\ref{fig:Eliashberg_anomali}(b) and (c). The
experimental gap values measured in anomalous contacts look to be
quite well reproduced by taking into account the reduction in the
$\sigma$-band DOS and the increase in interband scattering. The
interesting point is that the data in anomalous contacts
\emph{cannot} be fitted without interband scattering -- while the
data in standard contacts can, as clearly shown in
Fig.\ref{fig:Eliashberg_standard}. This is a consequence of the
different behaviour of the small gap in the two cases and suggests
that standard and anomalous contacts occur in regions of the sample
with different degrees of disorder.

\section{Possible origin of anomalous contacts}\label{sect:origin}
Let us summarize here the properties of anomalous contacts that
follow from the above. \\
1) Their critical temperature $T\ped{c}\apex{A}$ is much greater
than the bulk $T\ped{c}$ (from 10 to 25 K more).\\
2) In these contacts $\Delta\ped{\sigma}$ is very similar to that
measured in standard contacts with the same $T\ped{c}\apex{A}$,
while $\Delta\ped{\pi}$ is a little greater indicating a possible
enhancement of interband scattering. \\
3) The probability of finding anomalous contacts increases on
increasing neutron irradiation.

Properties (1) and (2) might indicate that anomalous contacts are
established in regions of the sample that are either less damaged or
partially ``reconstructed'', but in any case less disordered than
the surrounding material. Point (3), however, rules out the former
possibility so that anomalous contacts are most probably due to
reconstructed regions. It has been shown in a recent paper
\cite{Wilke} that heavily neutron-irradiated MgB$_2$ (with
$T\ped{c}$ as small as 5 K) thermally annealed at sufficiently high
temperatures (up to 500 $^{\circ}$C) and long times (24 h) almost
recovers all the characteristics (i.e. cell parameters, critical
temperature, residual resistivity) of the pristine samples. In our
case, there is no thermal treatment of the samples after
irradiation, so one can wonder whether similar annealing effects can
be due to the PCS measurements -- i.e., because of the current
locally injected in the sample through the point contacts -- or from
irradiation itself above a certain threshold dose.

Let us analyze first the hypothesis that local annealing occurs due
to the PCS technique. To do so, let us focus for convenience on the
contact made on sample P6 whose conductance curves are shown in
Fig.~\ref{Fig:raw_standard}. The normal-state resistance of this
contact is $R\ped{N}$ = 40 $\Omega$ and, as already said, the shape
of the curves (i.e. the presence of small dips \cite{dips} and the
offset of the curves on increasing temperature) tells us that the
contact must be in the diffusive regime \cite{Duif}. If only one
contact was established between sample and counterelectrode, its
radius $a$ -- evaluated from the resistance $R\ped{N}$ by means of
the Wexler formula (eq.~\ref{eq:Wexler}) -- would be $a \simeq 90$
{\AA}, which has to be compared with the effective mean free path
$\ell \simeq 5$ {\AA} and with the coherence length $\xi \simeq
100$~{\AA} \cite{Tarantini}. None of the two conditions for
spectroscopic analysis to be possible, i.e. $a \ll \ell$ (ballistic
conduction) and $a \ll \xi$, would be fulfilled. In this situation,
at bias voltages comparable to the energy gap, the carrier velocity
would largely exceed the depairing value \cite{Deutscher} and
superconductivity would be destroyed in a region close to the
contact of radius three times larger than $\xi$, with consequent
loss of the Andreev signal. Of course this contrasts with the
evidence of spectroscopic information present in the curves of
Fig.~\ref{Fig:raw_standard} (c). Moreover, in these conditions the
heating in the contact would be greater than experimentally
observed. In fact, using the standard equation valid for a circular
aperture \cite{Duif}
\begin{equation}
T\ped{max}^2= T\ped{bath}^2 + \frac{V^2}{4L} \label{eq:T}
\end{equation}
one obtains that, for a total voltage drop of the order of
$V\ped{peak}$, the temperature of the contact would reach the bulk
$T\ped{c}=8.7$~K when $T\ped{bath}=6.5$~K. This contrasts with the
curves reported in Fig.~\ref{Fig:raw_standard} where the difference
between $T\ped{c}$ and $T\ped{c}^{A}$ is less than 0.5~K.

To reconcile the experimental findings with the value of the contact
resistance, we are forced to admit that more than one contact is
established between sample and counterelectrode. Actually, this is a
rather natural assumption, considering the nature of our point
contacts whose macroscopic area is about 2000 $\mu$m$^2$. In our
case, it can be shown that the existence of $N\approx 20$ diffusive
contacts (supposed identical and with normal resistance $R\ped{N}$ =
800 $\Omega$) can perfectly explain the observed heating in the
contact region. In this case one obtains, \emph{for each contact},
$a$=8.2~{\AA}, and the temperature in the contact reaches the bulk
$T\ped{c}$ for $V=V\ped{peak}$ when $T\ped{bath}=8.2$~K. Moreover,
when the bias is of the order of magnitude of the gap, the current
density $j$ in the contact region is overcritical (take into account
that $j\ped{c} \simeq 2\cdot 10^5$ A/cm$^2$ in P6 \cite{Putti_APL})
but the distance $r\ped{j}$ in the superconductor over which $j$
decays to the critical value is of the order of 80 \AA $ < \xi$ thus
ensuring the spectroscopic properties of the contact.

\begin{figure}
\includegraphics[keepaspectratio, width=0.7\columnwidth]{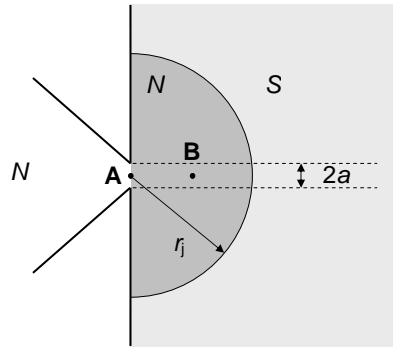}
\caption{Schematic representation of a point contact of radius $a$
with the region of radius $r\ped{j}$ driven normal by the current
density, which exceeds the critical value in the vicinity of the
striction. Points A and B are defined in the
text.}\label{fig:schema}
\end{figure}

It turns out from the above that the current injected \emph{during}
the measurement of the $I-V$ curve gives rise to little or no
heating in the contact region, even in the worst contact we
measured. However, it is worth recalling that, whenever necessary,
we tuned the normal-state resistance by applying voltage pulses of
some Volts for $\simeq 20 \div 80 $~ms, as experimentally
determined. At these bias values, $r\ped{j}$ is certainly greater
than $\xi$ so that a big normal region is formed (see
Fig.~\ref{fig:schema}) in which a very intense current flows for a
few tens of milliseconds. The normal region and the contact itself
are then quickly heated above the bath temperature
($T\ped{bath}=4.2$ K). It can be shown that temperatures of several
hundred Kelvin are easily reached in the contact. This is witnessed,
for example, by the early observation of an anomaly at about 250 mV
in the d$^2 V$/d$I^2$ of Fe-Fe homocontacts, associated with the
ferromagnetic transition of iron at the Curie temperature
$T\ped{C}=770$~K \cite{Verkin,Naidyuk}.

To evaluate the maximum temperature reached in our case, let $V$ be
the total measured voltage drop between sample and counterelectrode
and $V-V\ped{0}$ the voltage drop in the contact itself (let us
refer again to the contact on sample P6 for convenience), so that
$V\ped{0}$ is the potential difference across the (hemispherical)
normal region of radius $r\ped{j}$ (see Fig.~\ref{fig:schema}). If
$j\ped{0}$ is the current density in the orifice, then at a distance
$r$ from it one has
\begin{equation}
j(r)= j\ped{0}\cdot \frac{a^2}{2r^2}. \label{eq:j}
\end{equation}

The current flowing through the contact is $I =
(V-V\ped{0})/R\ped{PC}$. Here, $R\ped{PC}$ is the resistance of the
point contact according to eq.~\ref{eq:Wexler} where, in the Maxwell
term, $\rho(T)$ is the resistivity of the sample in the normal
state, at the temperature it will reach in the contact region
because of the Joule effect. Let us call $\rho\ped{ave}$ the average
of $\rho(T)$ in the temperature range to be determined (in sample P6
the RRR is so small that the value $\rho\ped{ave}\approx 160
\,\mu\Omega$cm can be acceptable for a wide range of temperatures).
Using these expressions one can calculate the voltage drop across
the normal region that is given by
\begin{eqnarray}
V\ped{0}& = & \frac{K\cdot V}{1+K} \text{             where} \nonumber \\
K & = & \frac{\rho\ped{ave}}{2 \pi
R\ped{PC}}\left(\frac{1}{a}-\frac{1}{r\ped{j}}\right). \label{eq:V0}
\end{eqnarray}
Using eq.~\ref{eq:j}, one also obtains
\begin{equation}
r\ped{j}= \sqrt{\frac{(V-V\ped{0})}{2 \pi R\ped{PC} j\ped{c}}}.
\label{eq:rj}
\end{equation}

In our case (20 contacts, each with $a=8.2$~{\AA}) and for $V=1$~V,
the solution of Eqs.~\ref{eq:V0} and \ref{eq:rj} gives
$V\ped{0}=0.257$~V and $r\ped{j}= 2576$~{\AA}. The maximum
temperature in the contact region (point A in
fig.~\ref{fig:schema}), evaluated from eq.~\ref{eq:T} and from the
voltage drop in the Maxwell part of the contact is $T\ped{A}\simeq
1300$~K. The maximum temperature reached in the normal region of
radius $r\ped{j}$ can be evaluated by asking that the thermal energy
generated within the normal volume by heating effects equals the
flux of heat current across the boundary. After suitably simplifying
the complex geometry of the problem, we estimate the maximum
temperature at the center of the normal region (point B in
fig.~\ref{fig:schema}) to be $T\ped{B} \simeq 640$~K. An alternative
approach to evaluate $T\ped{B}$ consists in using eq.~\ref{eq:T}
that, even if originally derived for a circular aperture
\cite{Duif}, approximately holds in this case too. This approach
gives the result $T\ped{B}\simeq 820$ K. These estimated
temperatures are higher than those used by Wilke et al.\cite{Wilke},
even if the duration of the heating process is by far shorter.
According to our results, annealing processes are most probable in
the contact region, i.e. close to the physical interface between the
two materials.

The hypothesis of local annealing as the origin of anomalous
contacts is thus very reasonable and well rooted in the physics of
point contact spectroscopy. However, we examined as well the other
possibility, i.e. pre-existent regions with higher $T\ped{c}$ than
the surrounding material, originated by local reconstruction due to
irradiation. Annealing effects due to the irradiation itself can
indeed occur, if irradiation takes place at low temperature, because
of the stimulated recombination of close Frenkel pairs
\cite{Averback}. If, otherwise, irradiation is carried out at room
temperature or above, competing phenomena such as creation and
annihilation of point defects or formation and coagulation of defect
clusters can partly compensate each other \cite{Kolontsova}, giving
rise to saturation in some physical parameters. Similar phenomena
are suggested, in our case, by the tendency to saturation in
$\rho\ped{0}$, $T\ped{c}$ and the $c$-axis parameter
\cite{Putti_APL} at very high neutron doses -- being the other
possible reason of saturation, i.e. the complete amorphization,
ruled out by the sharpness of the X-ray peaks. Whatever the exact
nature of the reconstruction process, locally-annealed regions
should feature higher critical temperature than the remaining part
of the sample, but since their presence is neither detected by
susceptibility, nor by specific heat and resistivity measurements,
they should represent a negligible part of the sample volume and
should be imagined as isolated regions of small size. Moreover, if
the macroscopic correlation between $\rho\ped{0}$ and $T\ped{c}$
observed in irradiated and annealed samples \cite{Putti_APL,Wilke}
is to be conserved also on a local scale, these regions are expected
to feature higher conductivity and greater density of states than
the surrounding matrix.

\begin{figure*}[t]
\includegraphics[keepaspectratio, width=\textwidth, angle=0]{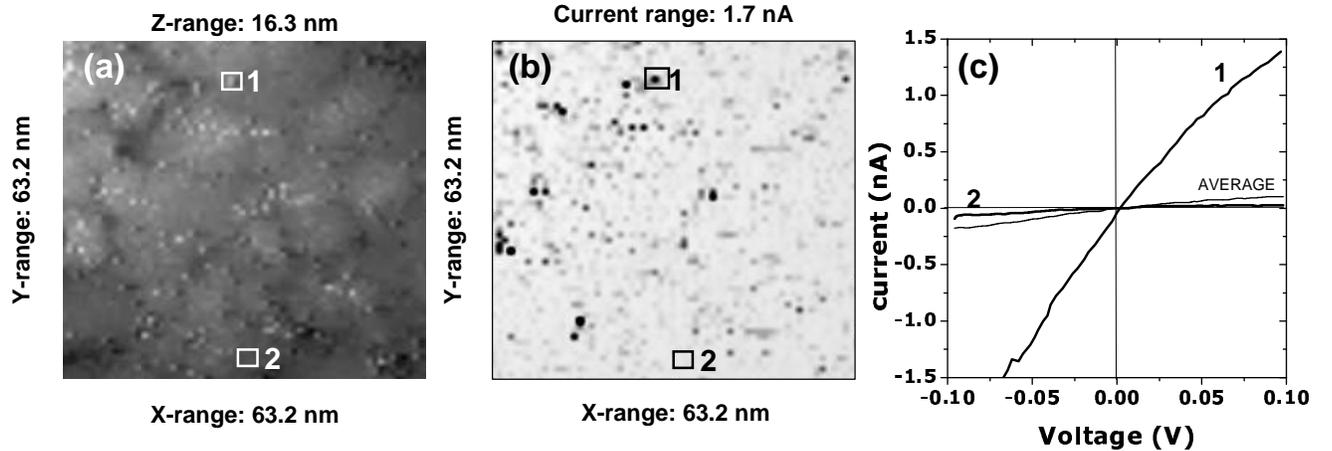}
\caption{(a) Scanning tunnel microscopy (STM) map of the surface of
a single grain in sample P6, measured at room temperature. The map
was taken with a fixed current equal to 2.0 nA. (b) Map of the
current across the N-I-N junction (tip/air gap/sample) measured at
constant tip-to-sample distance and constant voltage equal to 0.1 V.
(c) I-V characteristics of the N-I-N tunnel junction in points 1 and
2 of the topographical image (a). The average I-V characteristic is
also shown for comparison.}\label{Fig:sts}
\end{figure*}

With these ideas in mind, we performed room-temperature SEM and STM
analysis of the most irradiated sample, where the probability of
finding anomalous contacts was the highest. FESEM morphological
images of sample P6 showed large, well connected grains with smooth
surfaces. Microprobe analysis (EDX with SiLi detector sensitive to
light elements, B included) showed no trace of chemical species
other than Mg and B. Owing to the relatively large size of the
grains, we were able to perform scanning tunneling microscopy at
room temperature on the surface of a single grain. The topographical
image reported in Fig.~\ref{Fig:sts}(a) shows rather smooth
modulations on a length scale $\simeq 10$~nm in the $xy$ plane and
small brighter ``dots'' that look like protrusions. However,
morphological SEM images on the surface of grains on a similar scale
(100 nm $\times$ 100 nm) show no trace of such spots. The reason is
that STM is sensitive not only to morphology, but also to the local
density of states; regions with higher DOS appear brighter in the
STM map because they give rise to a higher conductance across the
tunnel junction. The topographical signal can be removed by
operating in STS mode, i.e. keeping the tip-to-sample distance
constant. The resulting map of the current measured across the
junction at a constant bias of 0.1 V is shown, in inverted gray
scale, in Fig.~\ref{Fig:sts}(b). The darker spots (corresponding to
higher currents) clearly correspond to the bright spots of
Fig.~\ref{Fig:sts}(a), while the smooth modulation is no longer
observed. For further confirmation, Fig.~\ref{Fig:sts}(c) reports
the I-V characteristics measured in two points (indicated by small
squares in panels (a) an (b)) together with the average of the I-V
curves measured in the whole region. It is clear that the ``dots''
observed in the STM maps represent very small regions (typically
$\varnothing \simeq 1$~nm) with higher conductivity with respect to
the surrounding material, that is exactly what one would expect for
locally annealed regions with higher critical temperature than the
bulk.

The problem now arises of understanding if these dots can be
superconducting above the critical temperature of the matrix in
spite of their very small size (if compared to the bulk coherence
length $\xi \simeq 10$ nm). A large number of experimental and
theoretical papers have been devoted to the so-called ``size
effect'' in isolated superconducting nanoparticles
\cite{Anderson,Muhlschlegel,Halperin,vonDelft} but very little is
known about the case in which these particles are embedded in a
conductive matrix. To the best of our knowledge, only one
experimental investigation was reported showing a reduction in the
critical temperature of lead nanoparticles embedded in a metallic
matrix on decreasing the particle size\cite{Tsai}. The reduction in
$T\ped{c}$ becomes effective below 20 nm (to be compared with the
bulk $\xi\simeq 90$ nm) but is much smoother than for Pb isolated
particles \cite{Li}, so that even for very small size ($\simeq 5$
nm) $T\ped{c}$ is only reduced by 30\% with respect to the bulk
value. On this basis, we can argue that regions of 3-5 nm in size
with a partially reconstructed lattice (which would be
superconducting even if isolated, according to Anderson's criterion
\cite{Anderson}) could well be superconducting even above the
critical temperature of the matrix and possibly give rise to
proximity effect on the surrounding normal material. Similar
regions, made up of clusters of nanoscopic ``dots'', are indeed
observed by STM in some part of the grain surface. Their density is
much lower than that of the bright dots in Fig.~\ref{Fig:sts}(a) and
this raises the problem of explaining the very high probability of
anomalous contacts. Again, the answer can be given by the technique
we used to tune the contact characteristics. Owing to their higher
conductivity with respect to the surrounding matrix, these regions
could in fact be privileged for the formation of new conduction
channels when a voltage pulse is applied. The resulting new contact
would then be dominated by the conductivity (and the critical
temperature) of these regions.

Both the proposed mechanisms of formation of anomalous contacts
require the application of a voltage pulse, either to provoke the
local annealing in the contact region or to select the preexistent
regions with higher conductivity (and possibly higher local
$T\ped{c}$). Looking for indirect support to our hypotheses, we
checked all the contacts we studied during several months and we
realized that indeed \emph{only ``modified'' contacts show anomalous
characteristics} and the few standard contacts we were able to
obtain in highly irradiated samples (among which the one shown in
Fig.~\ref{Fig:raw_standard}(c)), were actually ``as-made''. This
evidence further supports our picture and indicate that one of the
mechanisms described above could really explain the origin of the
anomalous contacts.

The local annealing induced by our PCS technique appears to be the
best understood and most likely process giving rise to anomalous
contacts. It easily accounts for all the experimental facts, i.e.:
i) the occurrence of anomalous contacts only after voltage pulses;
ii) the increasing-with-fluence probability to find such contacts
(related to the greater concentration of defects that can annihilate
on annealing); iii) the different behaviour of $\Delta\ped{\pi}$
with respect to standard contacts (due to the persistence of
additional disorder in the annealed regions, which is consistent
with the partial recovery of the pristine properties under annealing
\cite{Wilke}).

The second picture, in which anomalous contacts are established on
pre-existent regions partially reconstructed by irradiation requires
making some hypotheses: i) the regions with higher DOS observed by
STM are less disordered than the surrounding matrix; ii) regions of
3-5 nm in size can develop superconductivity even above the bulk
$T\ped{c}$; iii) the density of these regions is so small that the
probability for them to occur in "as-made" contacts is vanishingly
small; iv) these regions are privileged for the formation of new
conduction channels when a voltage pulse is applied. Some of these
points deserve further investigation, also for a better
understanding of the nature of defects in irradiated MgB$_2$.

Nevertheless, at this stage of investigation, it seems very unlikely
that the nanoscale inhomogeneities do not play any role in the
formation of the anomalous contacts; on the other hand, the
mechanism of local current-induced annealing appears very
convincing. The most reasonable picture is then that these two
effects coexist and interact. On one hand, local heating in the
contact region might help the migration of defects and their
clustering together with the partial re-arrangement of nuclei in the
lattice disordered by irradiation. On the other hand, more
conductive regions might really be preferred channels for the
current flow across the junction when the voltage pulse is applied,
and thus become the centers from which the annealing process starts.

\section{Conclusions}\label{sect:conclusions}
In conclusion, we have presented the results of point-contact
measurements in polycrystalline samples of neutron-irradiated
Mg$^{11}$B$_2$. By using the soft point-contact technique developed
for MgB$_2$ and related compounds, we measured the dependence of the
gaps on the local critical temperature of the contacts,
$T\ped{c}^{A}$. The resulting trend is in very good agreement with
the results of specific-heat measurements, especially in the most
irradiated samples, and perfectly confirms the first observation of
gap merging in undoped MgB$_2$ \cite{Putti_gap}. This is
particularly noticeable since the two techniques are completely
different and probe the surface and the bulk of the samples,
respectively. An analysis of the experimental gap trend within the
Eliashberg theory shows that a major role is probably played by the
decrease in the density of B $p\ped{xy}$ states, even if an increase
in interband scattering can be present as theoretically expected. A
fit of the gaps is however not possible in the whole range of
critical temperatures because in a certain intermediate region both
the gaps are smaller than the BCS value.

A striking experimental result was the occurrence of anomalous
contacts with $T\ped{c}^{A}$ higher than the bulk $T\ped{c}$, in a
percentage increasing with fluence and approaching 100\% in the most
irradiated sample. The conductance curves of these contacts are
perfectly fitted by the two-band BTK model, and their temperature
and magnetic-field dependencies are perfectly standard. However, the
trend of the small gap $\Delta\ped{\pi}$ extracted from their fit
differs from that obtained in standard contacts and can be
interpreted within the Eliashberg theory as being due to a more
effective interband scattering. Annealing effects, either due to our
particular PCS technique or to irradiation itself, have been
proposed to explain this anomaly. The two pictures have been
carefully investigated, both theoretically and experimentally. The
first one -- in which local annealing arises from the voltage pulses
we used to tune the contact resistance -- appears to be the most
likely, but the observation by STM of nanoscale regions with higher
DOS than the surrounding matrix could support the second as well.
Actually, a cooperative interaction of the two phenomena looks very
probable at this stage of investigation.

We are indebted to Ruggero Vaglio for suggestions about the
intepretation of anomalous contacts and to Pratap Raychaudhuri for
enlightening discussion about heating in the contact region. Special
thanks to Marina Putti for fruitful collaboration and help. This
work was done within the PRIN Project No. 2004022024 and the INTAS
Project No. 01-0617. V.A.S. acknowledges support by Russian
Foundation for Basic Research (Proj. No. 06-02-16490).

\end{document}